\documentclass[prl,twocolumn,showpacs]{revtex4}

\usepackage{epsfig}

\begin{document}
\newcommand{\ltwid}{\mathrel{\raise.3ex\hbox{$<$\kern-.75em\lower1ex\hbox{$\sim$}}}}
\newcommand{\gtwid}{\mathrel{\raise.3ex\hbox{$>$\kern-.75em\lower1ex\hbox{$\sim$}}}}
\newcommand{\bra}{\langle}
\newcommand{\ket}{\rangle}
\newcommand{\trace}{{\rm Tr}}
\newcommand{\ntilde}{\tilde{n}}
\newcommand{\stilde}{\tilde{s}}
\newcommand{\atilde}{\tilde{\alpha}}

\title{Finite Temperature Density Matrix Renormalization using an 
enlarged Hilbert space}

\author{Adrian E. Feiguin and Steven R.~White}

\affiliation{Department of Physics and Astronomy\\
University of California, Irvine, CA 92697}

\date{\today}

\begin{abstract}

We apply a generalization of the time-dependent DMRG to study 
finite temperature properties of several quantum spin chains, including 
the frustrated $J_1-J_2$ model. We discuss several practical issues
with the method, including use of quantum numbers and finite size effects. We
compare with transfer-matrix DMRG, finding that both methods produce excellent
results.  

\end{abstract}
\pacs{71.27.+a, 71.10.Pm, 72.15.Qm, 73.63.Kv}

\maketitle

The density matrix renormalization group (DMRG) method \cite{dmrg}
provides extremely accurate information about the ground state of
one dimensional systems. To study thermodynamic properties, it
was subsequently adapted to calculate the transfer-matrix of a 1D quantum
system. In the transfer-matrix DMRG (TM-DMRG) method \cite{tmdmrg,uri}, 
the usual DMRG sweeping takes place in
the imaginary time direction, whereas the thermodynamic limit in one
spatial direction is automatically obtained by targeting the maximum transfer
matrix eigenvalue and eigenvector. TM-DMRG gives excellent results, but is
also technically somewhat more difficult than ordinary DMRG, in part because the 
transfer matrix is non-Hermitian. A robust finite-temperature method based
on the original DMRG method would be very useful, if only because DMRG and TM-DMRG
have slightly different strengths and weaknesses.

In the past year, ideas from the quantum information 
field have greatly extended the capabilities of 
DMRG\cite{vidal1,vidal2,cirac1,cirac2}.
The first major advance was a new approach to real-time evolution within
a generalized DMRG framework (consisting of matrix-product 
states\cite{rommer}), and shortly
thereafter within a standard DMRG framework\cite{time, uri2}. 
Subsequently, separately Zwolak and Vidal, and Verstraete, et. al. devised
methods allowing finite temperature DMRG. The approach of Zwolak and Vidal
introduced the idea of a matrix-product description of density operators,
rather than of wavefunctions. Within the density operator formulation,
the infinite temperature system is trivial to describe, and imaginary time
evolution is used to reach finite temperature.
Verstraete {\it et al.} \cite{cirac-temp} argued that a more efficient
procedure is to enlarge the Hilbert space with auxiliary sites (called
ancillas), and to evolve in imaginary time a pure state within the 
larger space.
The auxiliary states act as a perfect heat bath, and when traced out
give exact thermodynamic averages. 

The ancilla approach is especially convenient from the traditional DMRG
point of view. 
A wavefunction description is usually more familiar and comfortable than
a density operator description. 
The ancillas appear geometrically as another chain parallel 
to the first, making the system resemble a ladder. More generally,
the ancillas form a copy of the original system, doubling the 
size of the lattice. 
As we discuss below, the ancilla states can be given quantum numbers, 
increasing
the efficiency of the calculation. In this paper, we apply the ancilla
approach to study the thermodynamics of several spin chains. We consider
practical issues such as finite size effects, and compare our results with
TM-DMRG.

The use of auxiliary systems to study thermodynamics in quantum systems 
originated as a key idea in thermo field dynamics 
\cite{thermofield,suzuki, barnett}. Let the energy eigenstates of the system in
question be $\{n\}$. Introduce an auxiliary set of fictitious states 
$\{\tilde{n}\}$ in one-to-one correspondence with $\{n\}$.
Define the unnormalized pure quantum state, in an enlarged Hilbert space,
\begin{equation}
| \psi(\beta) \rangle = e^{-\beta H/2}| \psi(0) \rangle = \sum_n 
e^{-\beta E_n/2} |n \tilde{n}\rangle \label{thermoa}
\end{equation} 
where $\tilde{n}$ is the matching state to $n$, $\beta$ is the inverse 
temperature, and $|\psi(0)\rangle=\sum_n{|n\tilde{n}\rangle}$ is our 
thermal vacuum. Then the partition function is 
\begin{equation}
Z(\beta) = \langle \psi | \psi \rangle
\end{equation}
and we can obtain the exact thermodynamic average of an operator $A$ (acting
only on the real states), as
\begin{equation}
\langle A \rangle = Z(\beta)^{-1} \langle \psi | A | \psi \rangle .
\label{average}
\end{equation} 

At $\beta=0$, the state $\psi$ is the maximally entangled state between
the real system and the fictitious system. If we change basis from the
energy eigenstates $n$ to some other arbitrary basis $s$, $\psi$ is 
still maximally
entangled\cite{suzuki}, $|\psi(0)\rangle = \sum_s |s \tilde{s}\rangle$. A natural 
basis to use is the site
basis, where the state of each site $i$ takes on a 
definite value $s_i$. One finds
\begin{equation}
| \psi(0) \rangle = \prod_i \sum_{s_i} |s_i \tilde{s}_i\rangle = \prod_i 
|I_i\rangle \label{site} \,
\end{equation} 
defining the maximally entangled state $|I_i\rangle$ of site $i$ with 
its ancilla.

>From the DMRG point of view, the maximally entangled state between the left
and right blocks would be the worst possible state to try to represent: all
density matrix eigenvalues would be equal. If one split the system between
the real sites on one side and the ancilla on the other, one would have 
exactly this worst case at $\beta=0$. 
It is remarkable that, if one pairs each site with its ancilla, 
and splits the system in two respecting this pairing, 
the infinite temperature state is the best possible state for DMRG, with only
one density matrix eigenvalue being nonzero. This leads to the natural ordering
of sites for DMRG site-1, ancilla-1, site-2, ancilla-2, etc. Alternatively,
one can group together a site and its ancilla into a supersite. Although the
dimension of the superblock is larger with supersites, next-nearest neighbor
interactions are not generated by the ancilla, simplifying the time evolution.
In our simulations we have used supersites.

The essence of the ancilla finite temperature method is to start in this
local $\beta=0$ state, and evolve in imaginary time through a succession
of temperatures $\beta$.
To evolve in time, we utilize one of the recently developed time evolution 
methods, which perform equally well in imaginary time.
The most efficient of these utilizes a 
Suzuki-Trotter break-up of the Hamiltonian, 
and each DMRG step consists of evolving the state using 
the link evolution 
operator $\exp(-\tau H_{i,i+1}/2)$ between the two central 
sites\cite{time, uri2}.
This method requires nearest neighbor interactions, at least in its 
simplest form.
Alternatively, one can evolve in a basis optimized for a single time 
step by solving explicitly the corresponding differential equation\cite{rk},
which does not require local interactions but is less efficient.
Note that the Hamiltonian only applies to the real sites; the ancillas 
evolve only through their entanglement.

In order to utilize conserved quantum numbers, it is useful to think of each
ancilla as being the anti-site of its site. A state of the ancilla is given 
opposite quantum numbers to the corresponding state of the real site. 
In this way, the state of interest has both total charge and total $z$ component of spin
equal to zero.

The infinite temperature starting state has a correlation length of 0 
and requires only one state per block.
As the system evolves in imaginary time, 
longer range entanglement is produced and the correlation length grows. 
The number of states needed for a given accuracy grows as the 
temperature decreases. It is most natural to slowly increase
the size of the basis, in order to keep a roughly constant truncation 
error. One may wish to set a minimum basis set size to make the early 
evolution essentially exact with little computational cost. In the test 
calculations below we kept the truncation error below $10^{-10}$, 
which in the systems considered corresponded typically
to a maximum of $m=500$ DMRG states. It turns out that most of these states
have $S^z=0$, and therefore the total size of the basis is of the order of
$10^6$ states for the spin-$\frac{1}{2}$ Heisenberg chain. 
In more difficult systems one would use a less stringent error criterion.

To illustrate the method, we begin by looking at the spin $S=1$ Heisenberg 
chain, using the Trotter time evolution method. 
We compare with the TM-DMRG results of Xiang\cite{Xiang S1}.
The TM-DMRG results have a small, well-controlled Trotter error owing to the 
formation of the transfer matrix; similarly, our time evolution has a different
small, well-controlled Trotter error. The TM-DMRG results are in the thermodynamic
limit, whereas our results here were on a $L=64$ site system with open boundary
conditions. 

\begin{figure}
\begin{center}
\epsfig{file=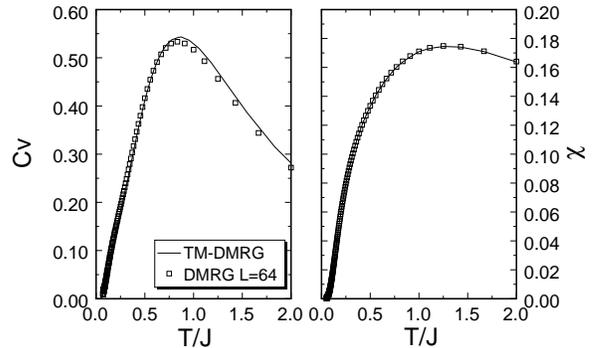,width=8cm}
\caption{
Specific heat and magnetic susceptibility of the $S=1$ spin chain of 
length $L=64$ obtained with the Suzuki-Trotter time-evolution algorithm. 
We compare with results from TM-DMRG in the thermodynamic limit.
}
\label{fig1}
\end{center}
\end{figure}

We calculated the specific heat $C_V$ by taking the numerical derivative of the
energy with respect to the temperature,  using energy differences between
adjacent time steps. In order to avoid edge effects we calculated the local energy in the center of the chain. 
We also calculated the magnetic susceptibility, using the formula
\[
\chi(T) = \frac{1}{T}\sum_{i}\langle S_0^z S_i^z \rangle,
\]
where the correlations were calculated at equidistant points from the 
center of the open chain, following the procedure described in 
Ref.\cite{huse}.
We have used half integer spins at both ends, as in Ref.\cite{bc}.
Results for these thermodynamic quantities 
are plotted in Fig.\ref{fig1}.
The agreement between the our results and Xiang's TM-DMRG is very good for both 
quantities, for temperatures down to $T\approx0.05$. 
At high $T$ we see slight deviations; we have checked our results at
high $T$ using smaller time steps and have found no difference, so we believe the
differences are due to Trotter error in the TM-DMRG.

As a second test example, we choose the spin-$\frac{1}{2}$ 
Heisenberg chain with nearest and next nearest neighbor interactions, 
with the Hamiltonian
\begin{equation}
H=\sum_i J_1 {\bf S}_i . {\bf S}_{i+1} + J_2 {\bf S}_i . {\bf S}_{i+2}.
\end{equation}
Since it is not trivial to use the Suzuki-Trotter break up for the 
frustrated case,
we used the time-step targeted method\cite{rk}.

\begin{figure}
\begin{center}
\epsfig{file=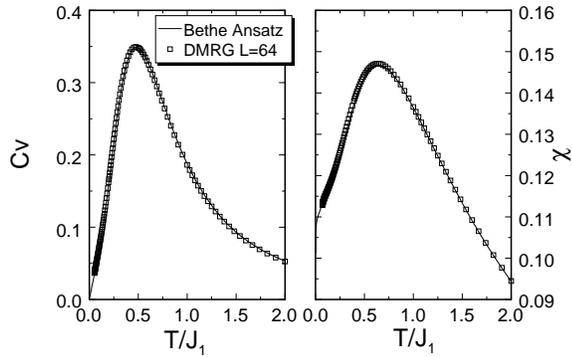,width=8cm}
\caption{
Specific heat and magnetic susceptibility of a $S=1/2$ spin chain of 
length $L=64$, compared to 
exact $L=\infty$ results using the Bethe Ansatz.
}
\label{fig2}
\end{center}
\end{figure}

In Fig.\ref{fig2} we compare our results for $C_V$ and $\chi$ 
for the unfrustrated chain ($J_2=0$) with 
results from the Bethe Ansatz calculations of Ref.\cite{bethe ansatz} in 
the thermodynamic limit. The agreement is excellent for the entire range of 
temperatures studied. Finite size effects were not apparent down to 
$T\approx0.1$.

When frustration is introduced, it is well known that this model is 
gapless for $J_2 \leq J_{2c} \approx 0.2411J_1$. At this value the chain 
breaks the translational symmetry by dimerizing, and an exponentially small 
gap opens.\cite{White and Affleck} At 
the point $J_2=0.5J_1$ the exact ground states become two dimer coverings, 
and the correlations extend only to one lattice spacing. The frustration 
present in this model makes reliable quantum Monte Carlo simulations very 
difficult, due to the appearance of the minus sign problem, and the most 
accurate results for thermodynamics quantities have been obtained using 
transfer-matrix DMRG. \cite{scholl, maeshima}

\begin{figure} \begin{center}
\epsfig{file=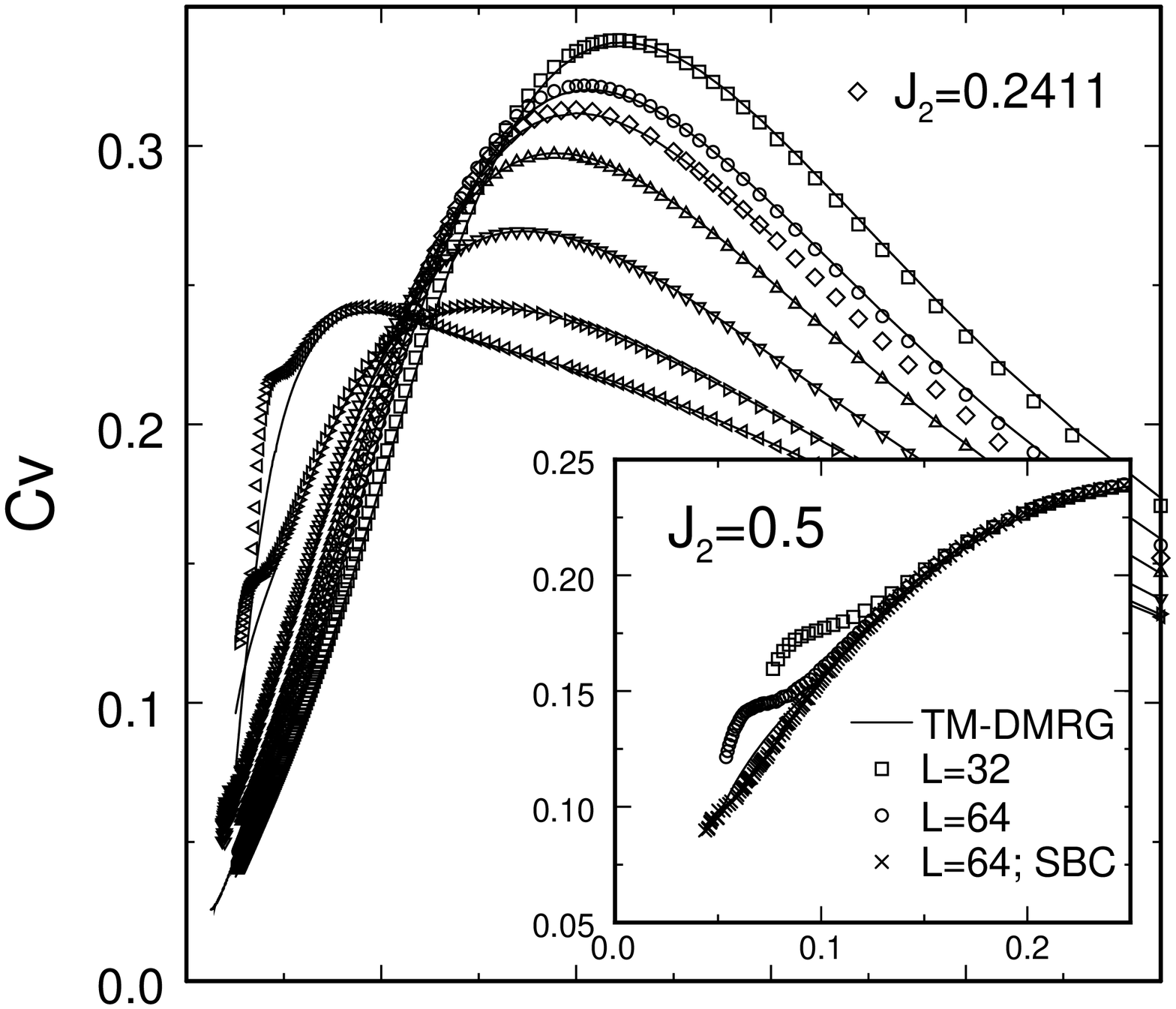,width=8cm}
\epsfig{file=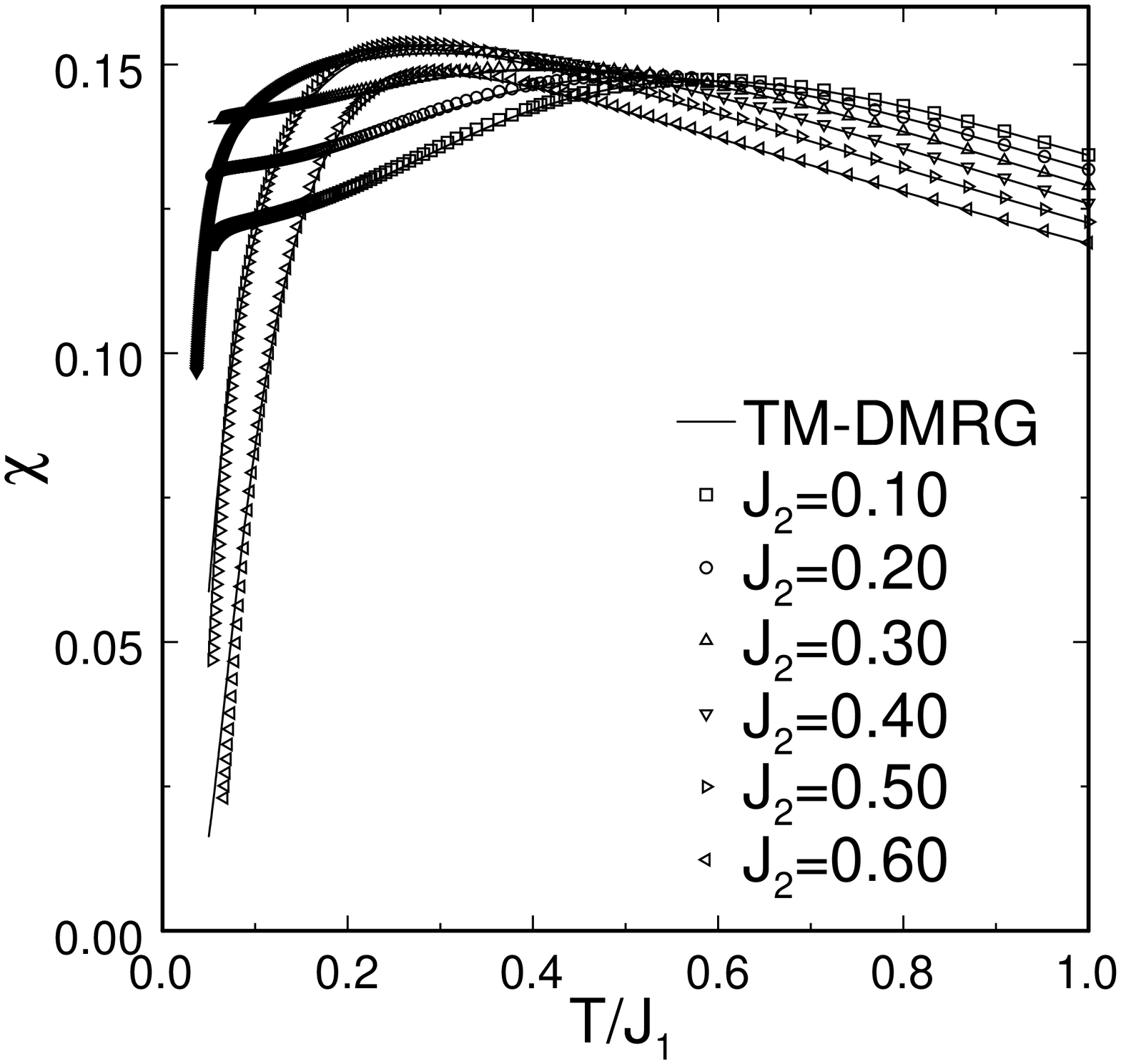,width=8cm}
\caption{
Specific heat and magnetic susceptibility of a frustrated $S=1/2$ spin 
chain of 
length $L=64$ obtained using time-step targeting, compared to 
results from transfer-matrix DMRG \cite{maeshima,scholl}. Inset: results for $J=0.5$ for different chain lengths and smooth boundary conditions.  
}
\label{fig3}
\end{center}
\end{figure}

In Fig.\ref{fig3} we show our results for the specific heat and 
susceptibility for different values of frustration $J_2$, below and 
above the critical point $J_{2c}$. Notice that due to dimerization, we 
have to symmetrize two correlations for each distance, $\langle 
S^z_{i}S^z_{j} \rangle$ and $\langle S^z_{i+1}S^z_{j+1} \rangle $. For 
small values of frustration, 
$J_2\leq J_{2c}$, the chain behaves as in the unfrustrated case, and the 
agreement is excellent, even for values of frustration up to $J_2=0.5J_1$. 
At this point we see more evident finite size effects reflected in the 
curves by a  
hump that appears in the specific heat at small temperatures. A detailed 
study of this hump shows that this is indeed a finite size effect, since 
it moves to lower temperatures as we increase the size of the chain as shown
in the inset in Fig.\ref{fig3}. At this value of the frustration, the 
ground state is degenerate in the thermodynamic limit, or in chain with 
periodic boundary conditions. This degeneracy is lifted in finite chains 
with open boundary conditions.
Similar characteristics can be observed in Ising chains, where the
two Neel configurations are ground states, when a small off-diagonal coupling
is introduced.\cite{bonner and fisher} 
By imposing a version of smooth boundary conditions \cite{smooth}, 
where we turn on $J_2$ slowly and smoothly from $0.0$ at the edges to $0.5J_1$ 
in the central region, we are able to eliminate the hump.

\begin{figure}
\begin{center}
\epsfig{file=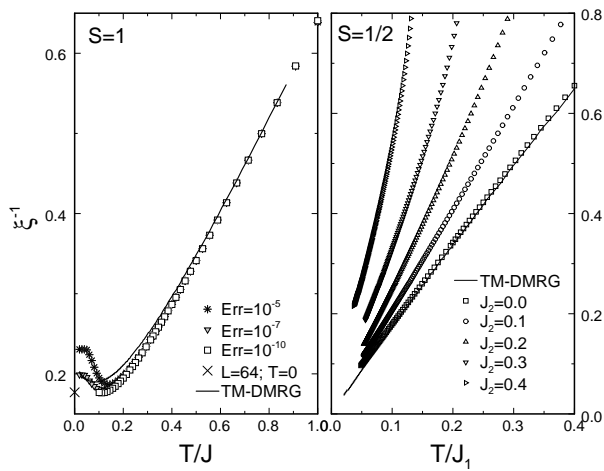,height=8cm,angle=-90}
\caption{
Correlation length as a function of the temperature for the $S=1$ Heisenberg chain (left panel), and the 
frustrated Heisenberg chain (right panel), 
 for different values of $J_2/J_1$. Our calculations are
on finite chains of length $L=64$.  
We add for comparison results from TM-DMRG in the thermodynamic limit.
For $S=1$ we show results using different truncation errors, and at $T=0$ from ground state DMRG.
}
\label{fig4}
\end{center}
\end{figure}

With our technique detailed spatial correlations functions are as easily obtained
as with ground state DMRG.
We have calculated the spin-spin correlations
and fit them to an expression of the form
\begin{equation}
C(r) \approx A \exp{(-r/\xi)}.
\end{equation}
Figure \ref{fig4} shows the results for $\xi^{-1}$ versus $T$
for $S=1$ and $S=1/2$.
The agreement with the results from TM-DMRG is excellent.  
For $S=1$ we notice the same minimum observed in the TM-DMRG simulation by Xiang \cite{Xiang S1}. We studied the system using time-targeting and also reducing
the time step, and we found that the source of that 
minimum can be attributed to the DMRG truncation error, as can be seen in the figure.

To summarize, we have described a DMRG algorithm to study strongly correlated 
quantum systems at finite temperatures by using an enlarged Hilbert space with 
ancillary degrees of freedom. We have illustrated its application by 
calculating thermodynamic quantities of gapless and gapped systems, including
frustration.
The ideas 
presented here are simple to implement as an extension to standard DMRG codes, 
and are not restricted to nearest-neighbor interactions or to single chains.

We thank J. Sirker for useful comments.   
We acknowledge the support of the NSF under grant DMR03-11843.


\end{document}